\documentclass[twocolumn,showpacs,preprintnumbers,amssymb]{revtex4}
\usepackage{graphicx}
\usepackage{dcolumn}
\usepackage{bm}

\begin{document}

\draft
\title{Directional interacting whispering gallery modes in coupled dielectric microdisks}
\author{Jung-Wan Ryu$^{1,2}$}
\author{Soo-Young Lee$^1$}
\author{Chil-Min Kim$^1$}
\author{Young-Jai Park$^2$}
\affiliation{$^1$ National Creative Research Initiative Center for Controlling Optical Chaos,\\
Pai-Chai University, Daejeon 302-735, Korea}
\affiliation{$^2$ Department of Physics, Sogang University, Seoul 121-742, Korea}

\begin{abstract}
We study the optical interaction in a coupled dielectric microdisks
by investigating the splitting of resonance positions of interacting
whispering gallery modes (WGMs) and
their pattern change, depending on the distance between the microdisks.
It is shown that the interaction between the WGMs
with odd parity about $y$-axis becomes appreciable at a distance
less than a wavelength and causes directional emissions of the resulting 
interacting WGMs.  The directionality of the interacting WGMs
can be understood in terms of an effective boundary deformation 
in ray dynamical analysis.
We also discuss about the oscillation of the splitting when the
distance is greater than a wavelength.
\end{abstract}


\maketitle
\narrowtext

\section{Introduction}

In spherical and cylindrical dielectric cavities high-$Q$ modes are the
whispering gallery modes (WGMs) in which light rays circulate along the 
curved inner boundary of the cavities, reflecting from the boundary 
with an incident angle always greater than the critical angle 
for total internal reflection, thus remaining
trapped inside the cavities\cite{McC92,Cha96}. 
There are only minute isotropic emissions of light
caused by evanescent leakage.
For the applications to optical communication and optoelectric circuit,
this isotropic emission is not desirable, rather directional emission
is much more useful and effective\cite{Cha96}.

As a simple system for directional emissions, slightly deformed
microcavities have been proposed, and directional emissions, tangential
from the boundary points with the highest curvature, are achieved.
In the ray dynamical viewpoint, as being slightly deformed, some invariant tori
are destroyed in the Poincar\'{e} surface of section (PSOS), but the Kolmogorov-Arnold-Moser (KAM) tori
still confine the rays supporting the WGMs.
In this case, the tunneling process, through the lowest dynamical barrier,
to the critical line for the total internal
reflection can explain the tangential emissions\cite{Cha96,Mek95}.

When the cavity boundary is highly deformed, the PSOS shows a global 
chaotic sea with very small integrable regions (islands).
In this strong chaotic case, the directional emissions can be found in scarred 
resonances\cite{Scar} and quasiscarred resonances\cite{Lee04a}.
Unlike the slightly deformed case, the direction of emission in these resonances
can be deviated from the tangential, and is well explained by the unstable
manifold structure near the critical line for total internal reflection\cite{Sch03}
and the Fresnel filtering effect\cite{Rex02}. 
In addition, there are special boundary shapes for generating unidirectional emission, 
spiral-shaped\cite{Spiral} and rounded triangle-shaped\cite{tri04}.
We note that the efforts for directional emissions are mainly based on
the deformation of boundary shapes.

In this paper we show, through a numerical study on the interacting WGMs
in a coupled identical disks,
that a mode-mode interaction can generate directional emissions.
The interaction between two WGMs is parameterized
by the distance between two disks, and it turns out that the strength of the
interaction between WGMs with odd parity about $y$-axis 
becomes appreciable at a distance less than a wavelength, which
are evident from the results on variation of resonance positions and patterns.
In order to explain the resulting directional emission, 
we assume that the circular boundary shapes would be effectively deformed due to 
the mode-mode interaction. With this assumption the ray dynamical analysis gives
a good explanation for the degradation of Q-factor and the enhancement of
directional emissions. In addition, when the WGMs are weakly coupled, the resonance
positions show an oscillating behavior depending on the distance.

The paper is organized as follows. In Sec. II we illustrate our system, i.e.,
a coupled dielectric microdisks.
The numerical results for the strongly interacting WGMs and a ray dynamical
model with an effective deformation are presented in Sec. III.
The behavior of weakly interacting WGMs is discussed in Sec. IV.
Finally, we summarize results in Sec. V.

\section{Coupled dielectric microdisks}

As a simple system for the study on interacting WGMs, we take coupled
dielectric microdisks. In this system, the strength of the interaction
can be controlled by adjusting the distance between two disks.
It is well known that the WGMs of a single dielectric microdisk are 
two-fold degenerate due to the circular symmetry, and are classified by 
the angular momentum mode index $m$ and the radial mode index $l$\cite{Cha96}.
Therefore, in the coupled microdisks, the WGMs would show
an approximate four-fold degeneracy when the distance $d$ between the disks
is very large, equivalently, the interaction between WGMs is negligible.
As the distance is getting smaller, the WGM in one microdisk starts to
know the existence of the WGM in the other microdisk, then the system is
no longer circular symmetric, and the four-fold degenerate resonance positions 
start to split each other, and the degree of the splitting measures the strength 
of the interaction between the WGMs.   

\begin{figure}[t]
\begin{center}
\includegraphics[width=0.4\textwidth]{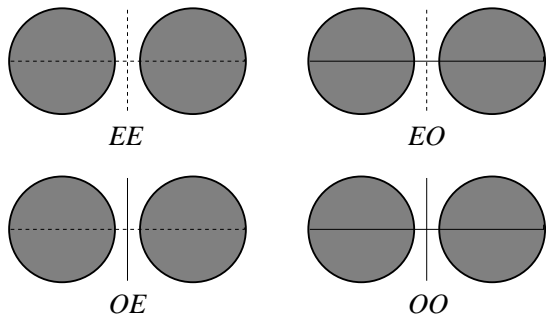}
\caption{$~$ The four symmetry classes of the coupled identical disks.
The former letter indicates the parity on $x$ and the latter does the parity on $y$
coordinates.
Even (Odd) symmetry is marked by dashed (solid) lines.}
\end{center}
\end{figure}

Figure 1 shows the four symmetry classes of coupled dielectric microdisks.
The system has two symmetry lines and the former letter $E(O)$ is even(odd)
if the wave function is even(odd) with respect to $x \rightarrow -x$
and the latter refers to $y \rightarrow -y$.

In practical calculation of the interacting WGMs in the coupled dielectric microdisks,
we use the boundary element method (BEM)
which is effective when the boundary is strongly deformed from a circular shape
and the cavities are coupled\cite{Wie03}.
In this paper, we focus on the TM polarization where both the wavefunction and its normal derivative
are continuous across the boundary.
The radius of disks is $R$, the distance between two dielectric disks is $d$,
and $n_{in}$ and  $n_{out}$ are the refractive indices inside and outside disks, respectively.  
We set $n_{out}= 1$ throughout the paper.

\section{Strongly interacting WGMs}

In this section we present numerical results on the variation of resonance
positions and patterns of the strongly interacting WGMs, i.e., the case of 
the short distance, $d < \lambda$. As mentioned before,
we expect that the four-fold degenerate WGMs would start to split
each other due to the interaction between WGMs as the distance $d$ decreases. 
As a result of the interaction the directional emissions appear 
in the interacting WGMs with odd parity about $y$-axis. 
We explain this directionality by assuming an effective deformation 
of boundary in ray dynamical analysis.
In the practical BEM calculation, we take 12 elements per a wavelength inside
($\lambda_{in}=2\pi/n_{in}k$).

\subsection{Variation of resonance positions and patterns}

Numerical calculation is performed for the WGM of mode index $(m,l)=(77,1)$
with $n_{in}=1.4$.
Exact resonance positions of the WGM in an isolated microdisk 
can be obtained from the matching conditions between the Bessel function 
and the Hankel function of the first kind which are inner and outer 
radial solutions of the Helmholtz equation, respectively. 
The exact resonance position of $WGM_{(77,1)}$ is
$kR=59.7136 - i2.5687\times 10^{-8}$ where $k$ is the vacuum wavenumber. 
The very small value of $ |\mbox{Im}(kR)|$ means high $Q$ factors 
from the relation $Q=-\mbox{Re}[kR]/2\mbox{Im}[kR]$. 
Unfortunately, it is very difficult to get the exact imaginary values from the BEM,
for example, when we take 12 elements per $\lambda_{in}$, 
the BEM calculation for the isolated microdisk gives 
$kR=59.7155 - i3.9\times 10^{-3}$ for $WGM_{(77,1)}$.
In spite of the restriction in the precision of resonance positions,
we rely on the BEM calculation in analyzing the interacting WGMs based on
the following reasons. First, the change of $\delta (kR)\simeq 10^{-3}$ does not
give any visible variation in resonance patterns. Second, the BEM calculation
would give a correct result when the variation of resonance positions exceeds
the precision limit, $\delta (kR)\simeq 10^{-3}$.

\begin{figure}[t]
\begin{center}
\includegraphics[width=0.4\textwidth]{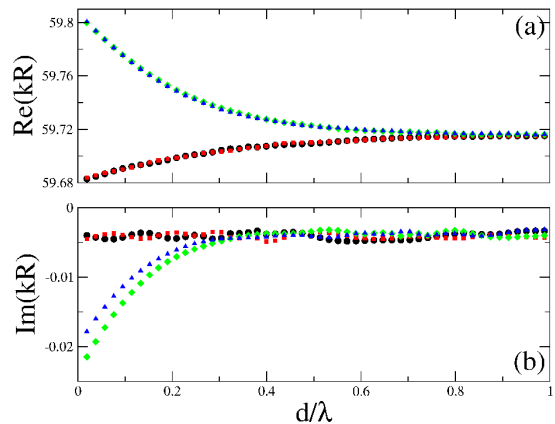}
\caption{$~$(color online) The splitting of the degenerate $WGM_{(77,1)}$
due to the mode-mode interaction in the coupled disks system.
$\mathrm{Re} (kR)$ (a) and  $\mathrm{Im} (kR)$ (b) are plotted depending on 
the distance $d$ between two disks. 
Note that $kR$ converges into the resonance position of 
the single disk case, $59.7155-i0.0039$ as  $d$ increases.
Black circle, red square, green diamond, and blue triangle correspond to
$EE$, $EO$, $OE$, and $OO$ modes, respectively.}
\end{center}
\end{figure}

Figure 2 shows the variation of resonance positions of the interacting
WGMs. The four fold degenerate state starts to split into two groups at $d\simeq 0.8\lambda$
in $\mbox{Re}[kR]$ and at $d\simeq 0.4\lambda$ in $\mbox{Im}[kR]$.
Here we can see that the resonances with odd parity about $y-$axis($OE$, $OO$ modes)  show
larger variations, indicating that the WGMs in the resonances are strongly coupled.
We note that the $\mbox{Re}[kR]$ values of $OE$, $OO$ modes increase with decreasing $d$.
The increment of $\mbox{Re}[kR]$ implies a reduction of effective boundary perimeter,
and the interaction between WGMs in the $OE$, $OO$ modes is, thus, repulsive.
From the same argument, we conclude that the interaction in the $EE$, $EO$ modes 
is weakly attractive. This result will be used in determining the effective 
deformation in the ray model in the next subsection.
As shown in Fig. 2 (b) the $\mbox{Im}[kR]$ values of $OE$, $OO$ modes
decrease with  decreasing $d$, implying the degradation of Q-factor 
due to the repulsive interaction. Therefore, we can expect that the emission
of the $OE$, $OO$ modes would be stronger than those of the $EE$, $EO$ modes.
From the viewpoint of the effective boundary deformation due to the repulsive
interaction, we can understand the difference of the onset points of the splitting
in $\mbox{Re}[kR]$ and $\mbox{Im}[kR]$.
In a slightly deformed cavity, the rays supporting the WGM are completely
confined by the KAM tori, resulting no drastic reduction of $Q$-factor.
As the cavity is more deformed, the KAM tori would be broken and the rays
can diffusively escape along the unstable manifolds, and then the $Q$-factor, 
equivalently $\mbox{Im}[kR]$, would decrease rapidly.

\begin{figure}[t]
\begin{center}
\includegraphics[width=0.4\textwidth]{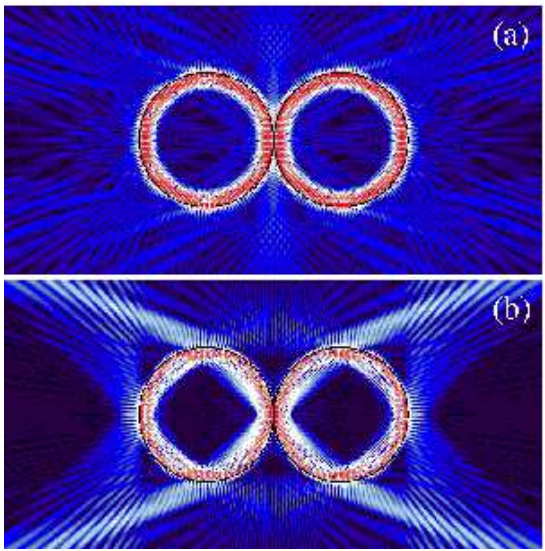}
\caption{$~$(color online) Resonance patterns of the interacting $WGM_{(77,1)}$s
when $n_{in}=1.4$ and $d=0.02\lambda$. (a) $EE$ mode.  (b) $OE$ mode.
Red-white-blue-dark blue colors indicate high to low intensity on a logarithmic scale.}
\end{center}
\end{figure}

From the splitting behavior of the resonance positions, we can expect that the
resonance patterns of $EE$ and $EO$ modes would be different from those of
$OE$ and $OO$ modes.
In Fig. 3 the resonance patterns of the interacting $WGM_{(77,1)}$s
are shown when the distance $d$ is $0.02\lambda$.
The resonance position of $EE$ mode shown in Fig. 3 (a) is
$kR=59.6829-i0.0040$ and,  as expected from the small absolute value of $\mbox{Im}[kR]$,
the very small evanescent leakage is shown\cite{com01}. 
However, as shown in Fig. 3 (b), the resonance pattern of $OE$ mode ($kR=59.7998-i0.0215$)
shows clear directional emissions, reflecting the strong repulsive mode-mode interaction.
Its far field emission pattern is plotted in Fig. 5 (a) where
the four strong directional emissions  are clearly seen.
We find that the directionality of the emission pattern is insensitive to the distance $d$,
although the strength of emissions decreases with increasing $d$.

\begin{figure}[t]
\begin{center}
\includegraphics[width=0.4\textwidth]{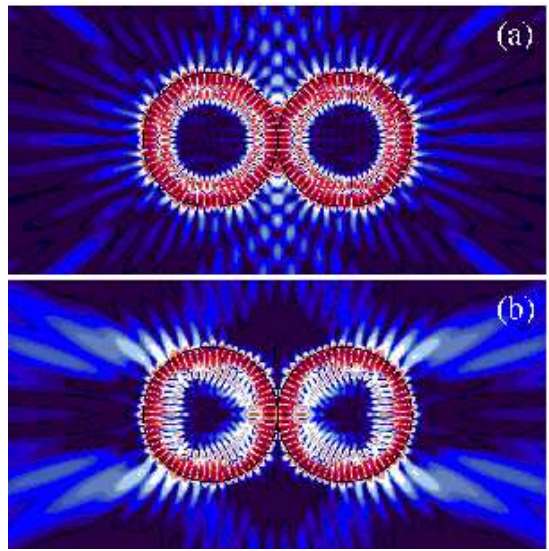}
\caption{$~$(color online) Resonance patterns of the interacting $WGM_{(29,1)}$s
when $n_{in}=2.0$ and $d=0.005\lambda$. (a) $EE$ mode.  (b) $OE$ mode.
Red-white-blue-dark blue colors indicate high to low intensity on a logarithmic scale.
}
\end{center}
\end{figure}

In $OE$ and $OO$ modes, the strong directional emission by the repulsive mode-mode interaction 
is a generic feature, but the emission directions are closely related to the reflective index $n_{in}$.
As an example, the resonance patterns of $EE$ and $OE$ modes in the interacting $WGM_{(29,1)}$s,
when $n_{in}=2.0$ and $d=0.005\lambda$, are shown in Fig. 4.
As expected, four strong directional emissions are shown only in the $OE$ resonance pattern 
in Fig. 4 (b),
and the two beams emitted from one disk are almost parallel to $x$-axis.
Note that the direction of emission is quite different from that of  the interacting $WGM_{(77,1)}$ in
Fig. 3 (b). The corresponding far field emission pattern is shown in Fig. 5 (b) where we confirm
the two directional emissions along $x$-axis and clear interference pattern of the parallel beams. 

In the next subsection we will introduce a ray dynamical model to explain the $n_{in}$ dependence
of the directionality of emissions in the interacting WGMs with $OE$ and $OO$ parities.

\begin{figure}[t]
\begin{center}
\includegraphics[width=0.4\textwidth]{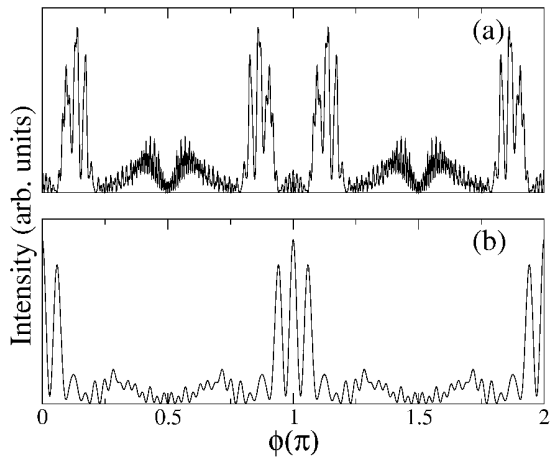}
\caption{$~$ The far field emission patterns of the interacting WGMs.
(a) The interacting $WGM_{(77,1)}$s ($n_{in}=1.4$) shown in Fig. 3 (b).
(b) The interacting $WGM_{(29,1)}$s ($n_{in}=2.0$) shown in Fig. 4 (b). 
}
\end{center}
\end{figure}

\subsection{Ray dynamical model : Effective deformation}

In a circular disk, the ray dynamics is simple. The rays with 
incident angles greater than the critical angle, $\theta_c=\arcsin (1/n_{in})$,
are perfectly confined in the disk by the total internal reflection,
while the other rays escape isotropically due to its rotational symmetry.
With a simple combination of this trivial ray dynamics, it is impossible to
explain the directionality of the interacting WGMs shown in the previous 
subsection.

We recall that the interaction between WGMs are repulsive in $OE$ and $OO$
modes. In fact, this originates from the constraint that the field value
at $x=0$ or on $y$-axis should be zero due to the odd parity about $y$-axis.
As a result, the intensity spots confronting each other  near $x=0$ would
shift repulsively, and the structure of the whole intensity spots in a WGM 
would be slightly deformed from circle.
In order to incorporate this effect of the repulsive interaction into
the ray dynamics, we consider a slightly deformed circular boundary which
can support the slightly deformed WGM similar to one of the interacting 
WGMs. 

\begin{figure}[t]
\begin{center}
\includegraphics[width=0.4\textwidth]{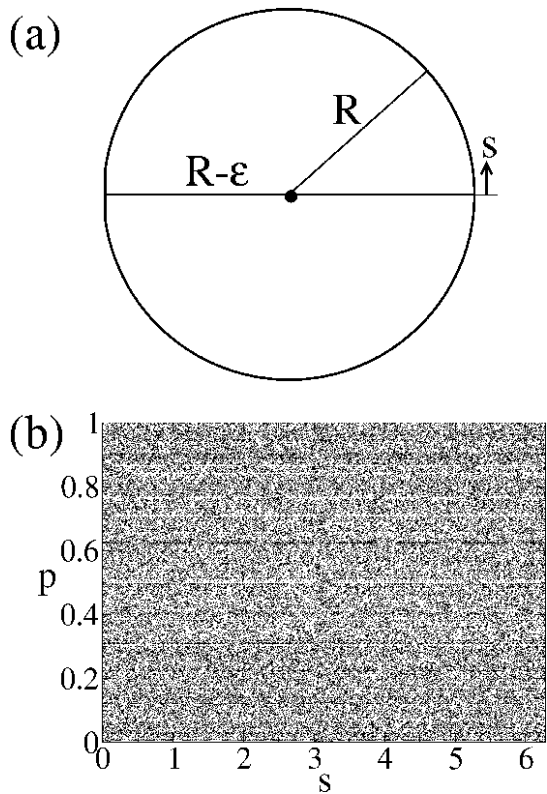}
\caption{$~$ (a) The boundary shape used in the ray dynamical analysis;
a circle with a cut.
(b) PSOS for the billiard shown in (a) when $R=1$ and $\epsilon=0.01$.}
\end{center}
\end{figure}

As a simple ray dynamical model containing the effect of the repulsive interaction,
we consider a circular disk with a cut as shown in Fig. 6 (a).
The deformation parameter is $\epsilon$ which is the reduced length of the radius
by the cut.
Figure 6 (b) shows its PSOS representing
the trajectory of a ray starting from one point in phase space $(s,p)$, where $s$ is the
boundary coordinate and $p=\sin \theta$, $\theta$ being the incident angle, 
without the consideration of the refractive escape.
Since the circle with a cut is a discontinuous deformation(non-KAM system),
this model cannot describe weakly deformed case where the rays supporting WGM
are still confined in KAM tori. So, our model is more suitable to a moderately deformed
case where the rays supporting WGM can diffusively escape.
The ray can change their incident angle only through the bounce on the cut,
and eventually the ray trajectory fill up the whole phase space, even though
there are so many marginally stable lines. 
The broken lines at $p \sim 0.50$ and $p \sim 0.71$ represent the families of the marginally
stable triangular and rectangular periodic orbits, and the gaps of lines correspond to 
the cut of the boundary.

\begin{figure}[t]
\begin{center}
\includegraphics[width=0.4\textwidth]{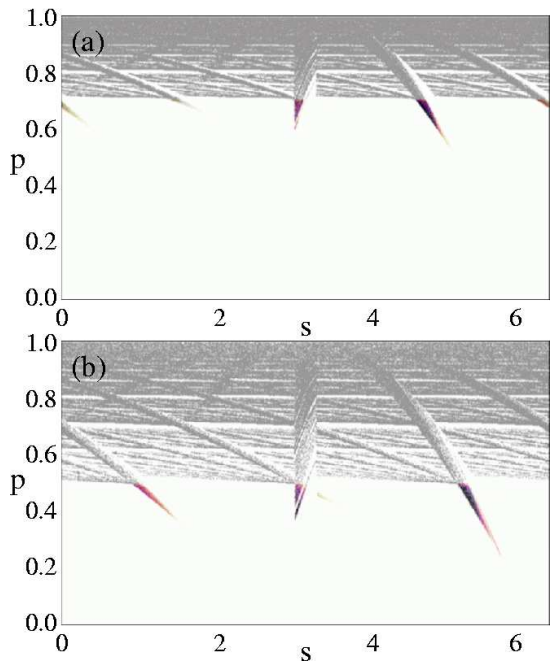}
\caption{$~$(color online) The survival probability distribution. 
The colored spikes show the parts through which long-lived rays escape.
(a) $n_{in}=1.4$ case.  (b) $n_{in}=2.0$ case.
}
\end{center}
\end{figure}

The dielectric microcavities are open systems where rays can refractively escape
from the microcavities, and the escape rate are determined by the Fresnel equations\cite{Haw95}.
In order to understand the emission direction of rays,
we obtain the survival probability distributions in both cases of $n_{in}=1.4$ and $n_{in}=2.0$
which are shown in Fig. 7 (a) and (b), respectively.
The gray points in Fig. 7 correspond to the rays with normalized probability 
greater than $0.1$ in the time range of $50 < t < 53$ with a time scale 
as the length of ray trajectory, and the rays start from a uniform ensemble 
of $1000 \times 1000$ initial positions in the phase space.
The pattern of the survival probability distribution reveals the openness structure
on the unstable manifold background near the critical lines, 
$p_{c} = 1/n_{in}$, for the total internal reflection.
The directionality of emissions and the emitting part of boundary are explained
by the pattern below the critical line.
The color plots below the critical lines in Fig. 7 (a) and (b)
illustrate how the long-lived rays supporting the WGMs can escape.
We find that the long-lived rays refractively escape through the unstable manifold
structure arising in the survival probability distribution. 
As mentioned before, the ray far above the critical line changes its angular momentum
only when bouncing from the cut and, depending on the bouncing position on the cut
the angular momentum can increases or decreases, in other words, the angular momentum
diffuses to other values. From repetition of this diffusion process,
the ray can reach below the critical line and eventually escape the microcavity.
The ensemble of the above escape process makes the color plots.
The darkness of the color plots represents the population of the ensemble.

The colored spikes below the critical line($p_c\simeq 0.714$) in Fig. 7(a) 
correspond to the rays following the diamond-typed period orbit inside the microcavity
and indicate that the ray emission from about $s=3\pi/2$ is very strong for the
counterclockwise circulating rays($p>0$).
These are consistent with the resonance pattern shown in Fig. 3 (b) where the faint
diamond structure is seen inside the microdisks and the strong emission comes out
from about $s=3\pi/2$ of the right microdisk.
If we consider another effective deformed microcavity to simulate the interacting WGMs,
we can get the resulting emission pattern shown in Fig. 8 (a) where we neglect the emission
from the cut. This ray dynamical result is very similar to the far field 
emission pattern of Fig. 5 (a) except the interference oscillation in peaks.
The same discussion is valid for the case of $n_{in}=2$ case shown in Fig. 7 (b).
The resulting emission pattern is given in Fig. 8 (b), and this explains well
the far field pattern of the interacting WGMs shown in Fig. 5 (b).

\begin{figure}[t]
\begin{center}
\includegraphics[width=0.4\textwidth]{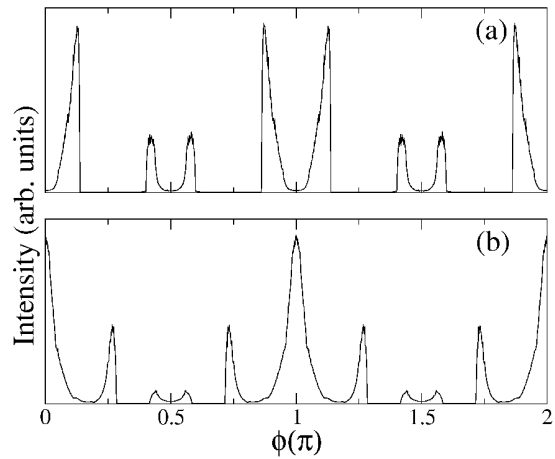}
\caption{$~$ The ray dynamical far field emission patterns of the deformed dielectric
disk. (a) $n_{in}=1.4$ case. (b) $n_{in}=2.0$ case.}
\end{center}
\end{figure}

Although the deformed disk in Fig. 6 (a) is a non-KAM system, this explains well 
the insensibility of the directionality of emissions to the distance $d$.
The longer distance $d$ corresponds to the smaller $\epsilon$ value.
In this case, although the average escape rate and the emitting part of the boundary
would decrease, it is clear that the emission directions 
and the emitting positions on the boundary are essentially invariant.
However, the mode-mode interaction would create a continuous deformation, i.e.,
the system would be a KAM system.
In a KAM model, even if the rays supporting the WGM are confined in
KAM-tori, it is still possible for the rays to reach chaotic sea through
the dynamical tunneling, and then diffuse along the unstable manifolds
to the critical line\cite{Pod05}. The unstable manifold structure near the critical line
in the KAM model would be similar to that of the non-KAM model if the global
boundary shapes of both models are almost identical. Therefore, both models
would give the essentially same emission directionality.

\section{Weakly interacting WGMs}

When the distance $d$ is larger than $\lambda$, the interaction between WGMs 
becomes very small. The strength of the small interaction can be measured
by the deviation of resonance position from that of isolated corresponding WGM as
done for the strongly interacting WGMs in the previous section.
Figure 9 (a) shows the variation of the resonance position in the range of 
$2\lambda < d < 6\lambda$ for the interacting $WGM_{(29,1)}$s with $n_{in}=2$.  
It is shown that the resonance positions of both $EE$(black circle) and $OE$(red square)
modes oscillate with the period $\Delta d \simeq \lambda/2$. The oscillatory behavior
 also appears for  $EO$ and $OO$ modes. The interference effect of emitted
waves from the WGMs seems to be responsible for the oscillatory behavior.
As pointed out in the previous section, the BEM calculation, however, has
the precision limit of $\delta(kR) \simeq 10^{-3}$. We note that the amplitude
of the oscillations is almost same order with the precision limit.
So, we have to be careful to accept the oscillatory behavior as a real phenomenon,
because it can be a numerical artifact. To check this, we performed the same
calculation for the interacting $WGM_{(7,2)}$s ($n_{in}=2.0$) which are 
relatively low-Q resonance modes. Since the emission of the $WGM_{(7,2)}$
is stronger than the $WGM_{(29,1)}$ case, we expect the variation of the resonance
position would be larger than the precision of BEM.

\begin{figure}
\begin{center}
\includegraphics[width=0.45\textwidth]{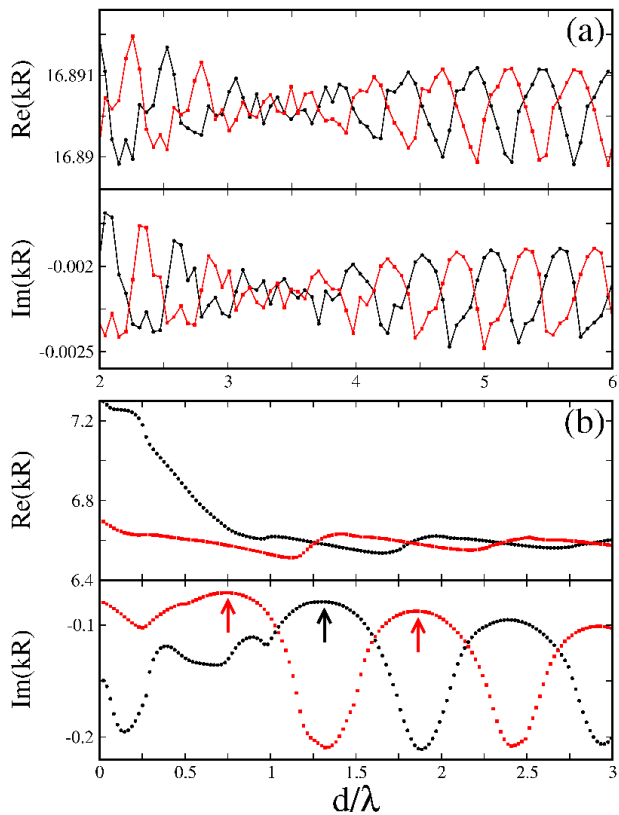}
\caption{$~$(color online) The oscillatory behavior of the resonance position of
interacting WGMs when  $d> \lambda$.
(a) $WGM_{(29,1)}$ case. 
(b) $WGM_{(7,2)}$ case.
Black circle and red square denote $EE$ and $OE$ modes, respectively.}
\end{center}
\end{figure}

The exact resonance position of the $WGM_{(7,2)}$ in an isolated circular disk 
is $kR=6.5806-i0.1117$ and in the BEM $kR=6.5808-i0.1127$ giving the
same precision limit of $\delta(kR) \simeq 10^{-3}$.
For the interacting $WGM_{(7,2)}$s, the variations of resonance positions for 
$EE$ and $OE$ modes  are shown in Fig. 9 (b).
When $d \gtrsim \lambda$, both real and imaginary parts of $kR$ oscillate with a period
$\Delta d \simeq \lambda$ and the oscillation amplitude is much greater
than the precision limit of the BEM, 
which supports that the oscillatory behavior of the interacting $WGM_{(29,1)}$ 
in Fig. 9 (a) would be a real phenomenon, not a numerical artifact.
We note that the oscillation of the resonance position of $OE$ mode
is out of phase with that of $EE$ mode in both $\mbox{Re}(kR)$ and $\mbox{Im}(kR)$.
The resonance patterns at the local maxima of $\mbox{Im}(kR)$, corresponding to minimum leakages
of the system, denoted by arrows in Fig. 9 (b) are shown in Fig. 10, and these explain why
the oscillations of resonance positions of  $OE$ and $EE$ modes are out of phase and
have a period  $\Delta d \simeq \lambda$.  
The internal patterns of the resonances are almost invariant, and the characteristic
difference is the number of intensity spots on the horizontal axis between two disks
which increases one by one. Odd and even number of spots appear in $OE$ and $EE$ modes,
respectively. Roughly we can understand this oscillation behavior as the degree of 
the accordance with the quantization condition of the unstable periodic orbit lying 
on the horizontal axis between two disks even though the boundary condition 
on the ends of the period orbit is not trivial.
In fact this kind of explanation is valid only for relatively low $kR$ case where
the interference on the unstable orbit dominates, and other inference effects are
negligible. In the high $kR$ case many beams emitted from the intensity spots inside
disks take part in the interference process, and the resulting oscillation of
each symmetry mode would show more complicated behavior in its period and amplitude.

\begin{figure}[t]
\begin{center}
\includegraphics[width=0.4\textwidth]{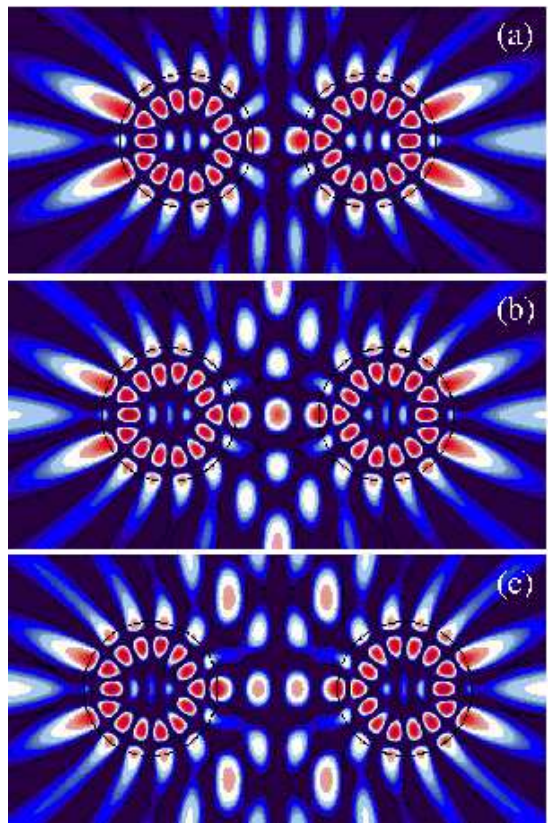}
\caption{$~$(color online) Resonance patterns of those corresponding to the
maxima of the oscillation indicated by arrows in Fig. 9 (b).
The resonance positions are (a) $kR=6.5744-i0.0712$ ($OE$ mode), 
(b) $kR=6.5824-i0.0793$ ($EE$ mode), and (c) $kR=6.5812-i0.0877$ ($OE$ mode).
$n_{in}=2.0$ is used and
red-white-blue-dark blue colors indicate high to low intensity on a logarithmic scale. 
}
\end{center}
\end{figure}

\section{Summary}

In the coupled disks system, we have shown that the strongly interacting WGMs 
with odd parity about $y$-axis gives good directional emissions,
and the directions of the emissions are determined by the
refractive index of the dielectric disks. This finding has been well explained
by an effective boundary deformation in ray dynamical model.
It is also shown that the resonance positions of the weakly interacting WGMs  
oscillate depending on the distance $d$ between two microdisks, and this oscillation
can be understood as the result of interference of emitted beams from the WGMs.

\section*{Acknowledgments}

This work is supported by Creative Research Initiatives of the Korean Ministry
of Science and Technology.


\begin{thebibliography}{150}

\bibitem{McC92} S. L. McCall, A. F. J. Levi, R. E. Slusher, S. J. Pearton, and R. A. Logan,
Appl. Phys. Lett. \textbf{60}, 20 (1992);Y. Yamamoto and R. E. Slusher, Physics Today \textbf{46}, 66 (1993).
\bibitem{Cha96} \textit{Optical Processes in Microcavities}, 
edited by R. K. Chang and A. J. Campillo (World Scientific, Singapore, 1996).
\bibitem{Mek95} A. Mekis, J. U. N\"ockel, G. Chen, A. D. Stone, and R. K. Chang, Phys. Rev. Lett.
\textbf{75}, 2682 (1995); J. U. N\"ockel, A. D. Stone, G. Chen, H. L. Grossman, and R. K. Chang,
Opt. Lett. \textbf{21}, 1609 (1996);
J. U. N\"ockel and A. D. Stone, Nature \textbf{385}, 45 (1997).
\bibitem{Scar}  E. J. Heller,  Phys. Rev. Lett. {\bf 53}, 1515 (1984);
S.-B. Lee, J.-H. Lee, J.-S. Chang, H.-J. Moon, S. W. Kim, and K. An,
Phys. Rev. Lett. {\bf 88}, 033903 (2002);
C. Gmachl, E. E. Narimanov, F. Capasso, J. N. Baillargeon, and A. Y. Cho,
Opt. Lett. {\bf 27}, 824 (2002);
T. Harayama, T. Fukushima, P. Davis, P. O. Vaccaro, T. Miyasaka,
T. Nishimura, and T. Aida, Phys. Rev. E {\bf 67}, 015207(R) (2003).
\bibitem{Lee04a} S.-Y. Lee, S. Rim, J.-W. Ryu, T.-Y. Kwon, M. Choi, and C.-M. Kim,
Phys. Rev. Lett. {\bf 93}, 164102 (2004).
\bibitem{Sch03} H. G. L. Schwefel, N. B. Rex, H. E. Tureci, R. K. Chang, A. D. Stone,
T. Ben-Messaoud, and J. Zyss, J. Opt. Soc. Am. B \textbf{21}, 923 (2004);
S.-Y. Lee, J.-W. Ryu, T.-Y. Kwon, S. Rim, and C.-M. Kim,
arXiv:nlin.CD/0505040 (2005).
\bibitem{Rex02}	N. B. Rex, H. E. Tureci, H. G. L. Schwefel, R. K. Chang, and A. D. Stone,
Phys. Rev. Lett. {\bf 88}, 094102 (2002);								
\bibitem{Spiral}G. D. Chern, H. E. Tureci, A. D. Stone, R. K. Chang, M. Kneissl, and N. M. Johnson,
Appl. Phys. Lett. {\bf 83}, 1710 (2003);
M. Kneissl, M. Teepe, N. Miyashita, N. M. Johnson, G. D. Chern, and R. K. Chang,
Appl. Phys. Lett. {\bf 84}, 2485 (2004);
T. Ben-Messaoud and J. Zyss, Appl. Phys. Lett. {\bf 86}, 241110 (2005).
\bibitem{tri04} M. S. Kurdoglyan, S.-Y. Lee, S. Rim, and C.-M. Kim, Opt. Lett. {\bf 29}, 2758 (2004).			
\bibitem{Wie03} J. Wiersig, J. Opt. A : Pure Appl. Opt. \textbf{5}, 53 (2003);
S.-Y. Lee, M. S. Kurdoglyan, S. Rim, and C.-M. Kim, Phys. Rev. A \textbf{70}, 023809 (2004).
\bibitem{com01} The intensity outside disks of Fig. 3 (a) is much smaller than that
of Fig. 3 (b) but for convenient sake, we show the near field intensity patterns with different
intensity scale. It is the same as Fig. 4. 								
\bibitem{Haw95} J. F. B. Hawkes and I. D. Latimer,
\textit{Lasers : Theory and Practice}, (Prentice Hall, 1995).
\bibitem{Pod05} V. A. Podolskiy and E. E. Narimanov, Opt. Lett. \textbf{30}, 474 (2005).

\end{thebibliography}
\end{document}